# INTEGRACIÓN DE PATRONES DE DISEÑO Y APLICACIONES MÓVILES EN UN SISTEMA DE GESTIÓN PARA EL CONTROL DE MANTENCIONES DE PLACAS CATÓDICAS DE LA CÍA. MINERA QUEBRADA BLANCA S.A.

# INTEGRATION OF DESIGN PATTERNS AND MOBILE APPLICATIONS IN A MANAGEMENT SYSTEM FOR MONITORING MAINTENANCE CATHODE PLATES OF MINING COMPANY QUEBRADA BLANCA SA.

Oscar Andrés Sandoval Carlos[1]     Wilson Andrés Castillo Rojas[2]

**RESUMEN**

Este documento presenta la integración de patrones de diseño y aplicaciones móviles, en el desarrollo de un software de gestión de placas (SIGEP) que permite apoyar en las soluciones a problemáticas que se presentan en el proceso de mantención de placas de cátodos de cobre de una Compañía Minera, en nuestro caso para Quebrada Blanca S.A. (CMQB S.A.). Estas problemáticas principalmente se relacionan con el escaso control sobre las tareas realizadas en las mantenciones a las placas catódicas, y la carencia de información que conlleva esta práctica, origina una gestión deficiente y no permite tomar decisiones oportunas referente a estos elementos, y por ende no permite proyectar y administrar la vida útil de las placas de cátodos, generando elevados costos asociados a este proceso. Al ser el proceso de mantención de placas de cátodo un proceso en constante cambio, con respecto a las estrategias de mantención, en el diseño del sistema SIGEP se valora la flexibilidad y reutilización en el diseño de los componentes del sistema, esto se logra gracias a los patrones de diseño utilizados. La implementación del sistema SIGEP y la incorporación de una aplicación móvil, significó para CMQB S.A. aumentar la fiscalización de las tareas realizadas a las placas de cátodos, permitiendo a la compañía contar con información detallada de las mantenciones de estos elementos, lo que permite entre otras cosas, identificar cuales placas de cátodo son más costosas par a la compañía y por tanto, conocer cuáles deben ser reemplazadas.

Palabras claves: Patrones de Diseño, Diseño Orientado a Objetos.

**ABSTRACT**

This document presents the integration of design patterns and mobile applications, in the development of software management of plates (SIGEP) that allows to support in the solutions to problematics that they appear in the process of maintaining of plates copper cathodes of a Mining Company, in our case for Quebrada Blanca S.A. (CMQB S.A.). These problematics mainly are related to the little control over the tasks carried out in the maintaining to the cathodic plates, and the lack of information that leads to this practice, originates a deficient management and it does not allow to make opportune decisions referring to these elements, and therefore it does to project and to administer the life utility of the plates of cathodes, generating lifted costs associated to this process.
As the process of maintaining a cathode plates constantly changing process, with respect to maintenance strategies in the system design SIGEP recognizing the flexibility and reuse in the design of system components, this achieved through design patterns used. The SIGEP implementation of the system and the incorporation of a mobile application, meant for CMQB S.A. increase control of the tasks carried out plates cathodes, allowing the company to detailed information on the maintenance of these elements, allowing among other things, identify cathode plates which are more expensive, and therefore knowing what must be replaced.

---

[1] Departamento de Ingeniería, Área de Computación e Informática. Universidad Arturo Prat. Av. Arturo Prat 2120. Iquique, Chile. E-mail: osandoval.inf@gmail.com.

[2] Departamento de Ingeniería, Área de Computación e Informática. Universidad Arturo Prat. Av. Arturo Prat 2120. Iquique, Chile. E-mail: wilson.castillo@unap.cl.

## INTRODUCCIÓN

Actualmente el proceso productivo de la planta de electrobtención de la Compañía Minera Quebrada Blanca S.A. (CMQB S.A.) trabaja con placas de cátodos de cobre, las cuales a través de un proceso de diferencial de potencial eléctrico, logran la adhesión de las moléculas de cobre que se encuentran en la solución en la cual están sumergidos.

La Superintendencia de Planta de CMQB S.A., cuenta con alrededor de 16.000 placas catódicas en operación, las cuales se someten a mantenciones diarias y mensuales, éstas consisten en el pulido y limpieza de las placas de cátodos de cobre. Las placas catódicas a lo largo de su vida útil, muestran variaciones de costos respecto a sus mantenciones, debido a una infinidad de variables, como la calidad de la mantención o el deterioro de éstas. Las placas de cátodos que han excedido su vida útil, son más costosas para la compañía, debido a que requieren ingresar a mantención con mayor frecuencia, por lo tanto, identificar estas placas de cátodos y reemplazarlas por placas nuevas será muy útil para reducir el costo por este concepto.

Asimismo, llevar un adecuado historial y control de estas mantenciones permitirán a la compañía entre otras cosas, anticiparse al término de la vida útil de la placa catódica, conocer que placas demandan mayor costo de mantención y por lo tanto, conocer cuáles deben ser reemplazadas, lo que ahorrará a la compañía tiempo y dinero por concepto de mantenciones de este elemento. Contar con este software que cubra esta necesidad de información, le ha permitido a CMQB S.A., tener información clara, persistente, oportuna y automática para tomar correctas decisiones relacionadas con las mantenciones de las placas catódicas.

Se realizan en promedio entre 200 y 250 mantenciones de placas catódicas por día, para asegurar el correcto funcionamiento de este proceso, el operador fiscaliza en terreno estas mantenciones, para ello requiere registrar la información de estos trabajos al instante en que se realiza esta actividad, por este motivo se decide integrar una interfaz móvil del software que permita al operador, realizar la toma de datos en línea en la misma faena, y así asegurar una correcta supervisión de las tareas realizadas a las placas de cátodos de cobre por concepto de mantenciones. El proceso que se lleva a cabo en faena para la mantención de las placas catódicas, comienza cuando el operador placa decide, de acuerdo al estado de las placas, que tipo de tareas de mantención necesita, mantención mayor, para placas muy deterioradas y mantención menor, para placas con menor deterioro. Las placas de cátodo reparadas se almacenan en un stock de placas disponibles a la espera de ingresar a producción, se puede representar este proceso, mediante el siguiente diagrama que muestra la figura 1:

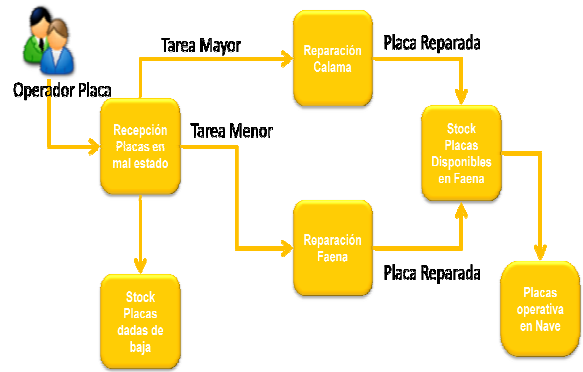

Figura 1. Diagrama del proceso de gestión de mantenciones placas catódicas.

El desarrollo de este software de Gestión de Mantenciones de Placas Catódicas, de ahora en adelante SIGEP, se basa principalmente en el diseño orientado a objetos bajo la asignación de responsabilidades. Uno de los objetivos que se buscan al utilizar esta técnica es conseguir la reutilización de software.

Para esto, se han identificado y utilizado patrones de diseño, que hoy son un mecanismo efectivo de reutilización, y se han convertido en una técnica popular para el re-uso de conocimiento en el diseño de software,

Adicionalmente, para apoyar y agilizar el proceso de control y gestión de mantenciones de las placas de cátodos de cobre en terreno, se han integrado herramientas móviles para facilitar la toma de datos en línea dentro de la faena misma.

## MARCO TEÓRICO

La orientación a objetos, es un enfoque de desarrollo de software que organiza tanto el problema como su solución, en una colección de objetos discretos, tanto la estructura de datos como el comportamiento están incluidos en la representación [1].

Una representación orientada a objetos puede reconocerse por sus siete características: identidad, abstracción, clasificación, encapsulamiento, herencia, polimorfismo y persistencia. Algunas representaciones sólo utilizan un subconjunto compuesto por parte de estas características, estas representaciones casi orientadas a objetos suelen denominarse basadas en objetos.

Se debe distinguir que la Programación Orientada a Objetos (POO) como paradigma, enfoque o manera de visualizar la realidad, y como metodología, colección de características para la ingeniería de software, no son las mismas cosas. Se puede lograr una notoria diferencia a través de los conceptos para cada uno de estos puntos de vista [2].

Dentro del desarrollo de cualquier tipo de software, es necesario tener un análisis y diseño bien definido, pero sobre todo una correcta arquitectura de software. El término arquitectura de software ha sido definido de diferentes maneras, pero muchos autores coinciden con la siguiente definición: "es una estructura compuesta de componentes, conectores, y reglas que definen como combinarlos, caracterizando la interacción de esos componentes" [3].

Se debe considerar que todo problema se puede caracterizar por ciertos patrones a seguir, los mismos que corresponden a una posible solución, y que al combinarse, pueden producir cierta solución a cualquier problema establecido. En la resolución de diversos tipos de problemas de software se han encontrado dichos patrones durante el proceso de diseño, motivo por el cual, han sido llamados patrones de diseño. Estos patrones hacen que el diseño sea más flexible, elegante y extremadamente reutilizable, por los que hacen crecer las posibilidades de encontrar una solución de manera rápida.

Para describir claramente un patrón de diseño, se debe hacer de una manera detallada y con un formato consistente, cada patrón es dividido en secciones de acuerdo a una plantilla [4]. Esta plantilla proporciona una estructura uniforme a la información, de tal manera que el patrón de diseño es sencillo de entender, comparar y utilizar. Cada una de estas especificaciones es importante debido a que el principal motivo de los patrones de diseño es la reutilización, por lo que su existencia no tiene sentido si es que no son aplicables en este campo.

Los patrones como elementos de reutilización tienen sus principios en la arquitectura (en el sentido clásico), donde fueron introducidos con el objetivo de capturar y posteriormente utilizar diseños que se habían aplicado en otras construcciones y que se catalogaron como complejos. El arquitecto Christopher Alexander [5], fue el primero en tratar de crear un formato específico para patrones en la arquitectura. Argumentaba que los métodos comunes aplicados en la disciplina daban lugar a productos que no satisfacían las demandas y requerimientos de los usuarios, y que eran ineficientes a la hora de conseguir el propósito de todo diseño y esfuerzo de la ingeniería: mejorar la condición humana. Describió algunos diseños eternos para tratar de conseguir sus metas. Propuso así, un paradigma para la arquitectura basado en tres conceptos: la calidad, la puerta y el camino.

C. Alexander formuló la siguiente definición de Patrón:

*"Cada patrón describe un problema que ocurre una y otra vez en nuestro entorno, para describir después el núcleo de la solución a ese problema, de tal manera que esa solución pueda ser usada más de un millón de veces sin hacerlo ni siquiera en dos oportunidades de la misma forma"* [5].

En el marco de la ingeniería de software, existen diferentes ámbitos donde se pueden aplicar patrones. La siguiente es una de las posibles clasificaciones: Patrones de Arquitectura, Patrones de Diseño, Patrones de Análisis, Patrones de Proceso o de Organización, y los denominados Idiomas. La diferencia entre estas clases de patrones está en los diferentes niveles de abstracción y detalle, y del contexto particular en el cual se aplican o de la etapa en el proceso de desarrollo.

Un patrón de este tipo identifica, abstrae y nombra los aspectos elementales de una estructura de diseño, donde los componentes, son las clases y objetos, y sus mecanismos de interacción son mensajes. Cada patrón de diseño especifica una estructura de clases, roles y colaboraciones, y una adecuada asignación de métodos para resolver un problema de diseño en una manera flexible y adaptable [3].

Uno de los beneficios de utilizar patrones es el entendimiento y documentación de diseños orientados a objetos. Los patrones de diseño, también, mejoran el mantenimiento de sistemas ya que proveen una especificación explícita de clases e interacción entre objetos. Además, los patrones de diseño proveen un vocabulario para discutir y comunicar decisiones de diseño en término de estructuras de clases en lugar de objetos.

GRASP es el acrónimo de General Responsibility Asignment Software Patterns (patrones generales de software para asignar responsabilidades). El nombre se eligió para indicar la importancia de captar estos principios, si se quiere diseñar eficazmente el software orientado a objetos. Los patrones GRASP constituyen un apoyo para la enseñanza que ayuda a entender el diseño de objetos esencial, y aplica el razonamiento para el diseño de una forma sistemática, racional y explicable. Este enfoque para la comprensión y utilización de los principios de diseño se basa en los patrones de asignación de responsabilidades [6].

Los patrones GRASP no establecen nuevas ideas: son una codificación de principios básicos ampliamente utilizados. Por mencionar algunos de los principales patrones GRASP, tenemos: Experto en Información, Creador, Alta Cohesión, Bajo Acoplamiento, y Controlador [6].

Y cuatro patrones adicionales que son: Polimorfismo, Indirección, Fabricación Pura, Variaciones Protegidas.

Por otro lado, los patrones GoF (acrónimo Gang Of Four), comúnmente conocidos como la "Banda de los Cuatro", y que corresponde a los cuatro autores del libro Design Patters [4], y que en 1995 plasman los conceptos de C. Alexander en este libro, su aporte es importante ya que estableció los requisitos que debe poseer un patrón de diseño orientado a objetos, y recopiló los patrones de diseño básicos.

Definieron 23 patrones de diseño, y los clasificaron en tres categorías:

i) Creacionales: los cuales tratan con las formas de crear instancias de objetos. El objetivo de estos patrones es de abstraer el proceso de instanciación y ocultar los detalles de cómo los objetos son creados o inicializados.
ii) Estructurales: describen como las clases y objetos pueden ser combinados para formar grandes estructuras y proporcionar nuevas funcionalidades. Estos objetos adicionados pueden ser incluso objetos simples u objetos compuestos.
iii) Comportamiento: ayudan a definir la comunicación e iteración entre los objetos de un sistema. El propósito de este patrón es reducir el acoplamiento entre los objetos.

## DISEÑO DE SIGEP CON PATRONES GRASP

La siguiente sección muestra las elecciones y decisiones tomadas durante el diseño para la realización de los casos de usos con objetos basado en los patrones GRASP [5].

Lo esencial de un diseño de objetos lo constituye el diseño de las interacciones de objetos y la asignación de responsabilidades. Las decisiones que se tomen pueden influir profundamente en la extensibilidad, claridad y mantenimiento del sistema de software de objetos, además en el grado y calidad de los componentes reutilizables.

El modelo de dominio, no representa clases de software, pero se puede utilizar para inspirar la presencia y los nombres de algunas clases de software en el modelo de diseño. Por lo tanto, se crea un diseño con una representación más baja entre el diseño del software y la percepción del dominio del mundo real con el que el software está relacionado.

Para crear un modelo de dominio de clases conceptuales interesantes o significativas del dominio de interés, en este caso la gestión de placas de cátodos de cobre, la tarea central es, identificar las clases conceptuales relacionadas con el escenario que se está diseñando. A continuación se presenta una lista de categorías de clases conceptuales, la lista está restringida al escenario simplificado de gestionar placas de cátodos y gestionar mantenciones de placas de cátodos.

- Placas
- Mantención
- Tareas
- Estados
- Empresa
- Operador Placa

El modelo de dominio, representado en la figura 2, muestra a Operador Placa, como responsable de gestionar las placas de cátodos, iniciar una mantención y designar el tipo de tareas a realizar en la mantención de las placas de cátodos.

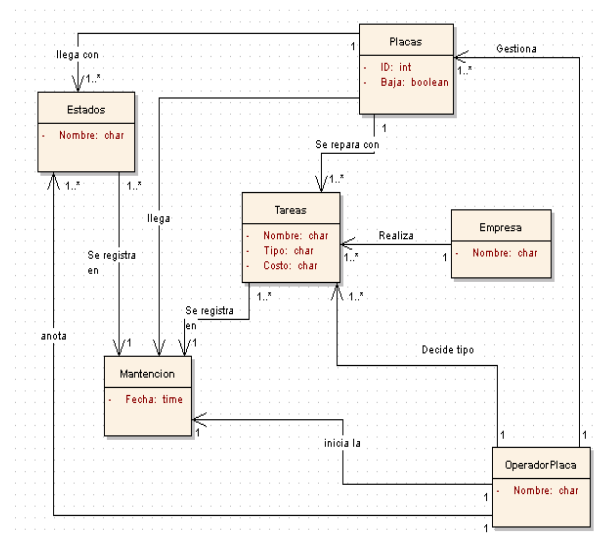

Figura 2. Modelo de Dominio Gestión Mantenciones Placas

Dado este modelo de dominio restringido al proceso de mantención de placas de cátodos, se identificaron dos casos de usos principales, *Gestionar Placas y Gestionar Mantenciones*, tal como se puede observar en el siguiente diagrama de casos de uso.

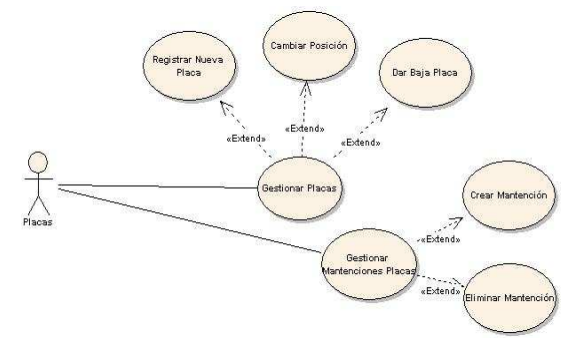

Figura 3. Diagrama de Casos de uso

La primera elección de diseño comprende la elección del controlador para los mensajes de operación del sistema *RegistrarNuevaPlaca, Cambiar Posición, Dar Baja Placa, Crear Mantención y Eliminar Mantención*.

Es aceptable elegir un controlador de fachada como una única clase, si sólo hay pocas operaciones del sistema y el controlador no está asumiendo demasiadas responsabilidades, en otras palabras, si va a perder la cohesión. Es adecuado elegir un controlador de caso de uso cuando hay muchas operaciones del sistema y deseamos distribuir las responsabilidades con el fin de mantener a cada clase controladora ligera y centrada, es decir cohesiva. En este caso, se utilizará una clase, puesto que sólo hay unas pocas operaciones del sistema.

Consultando el modelo de dominio, *OperadorPlaca* es un buen candidato para ser la clase controladora. Por tanto, el diagrama de interacción que se muestra en la figura 4 comienza enviando los mensajes *Registrar Nueva Placa*, *Cambio Estado*, *Dar Baja Placa*, *Crear Mantención y Eliminar Mantención* al objeto software *OperadorPlaca*.

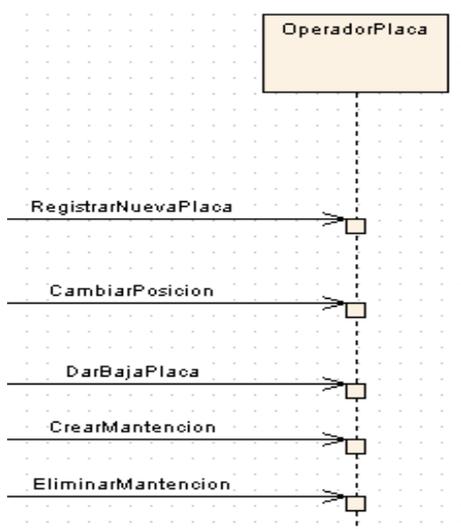

Figura 4. Aplicación del patrón GRASP Controlador.

Para el diseño de objetos, se debe crear un nuevo objeto software Placa, y el patrón GRASP Creador sugiere la asignación de la responsabilidad de creación a la clase que agrega, contiene o registra el objeto que se va a crear.

Por tanto, el *OperadorPlaca* es un candidato razonable para crear un objeto Placa. Según el mundo real el Operador Placa es el responsable de registrar, dar de baja y cambiar la posición de un placa de cátodo de cobre, por tanto es un experto en información en estas actividades y sería razonable que el objeto software *OperadorPlaca* fuera el responsable de estas funcionalidades.

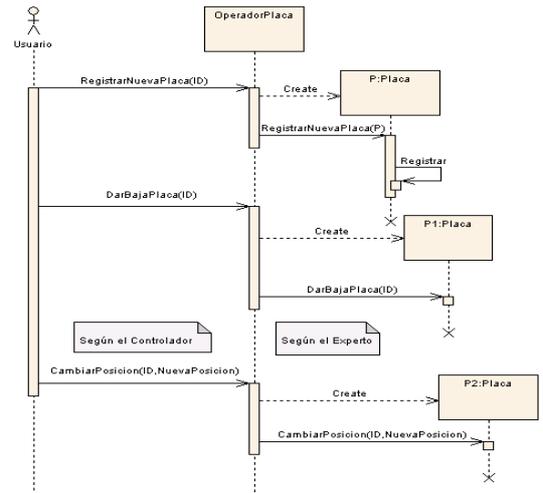

**Figura 5.** Diagrama de Secuencia Gestionar Placas

Para *Gestionar Mantención*, se asume la necesidad de crear las instancias: Placas, Tareas y Estados y asociarlas a Mantención. ¿Qué clase debería ser responsable de esto?. Puesto que un operador placa registra una nueva mantención en el domino del mundo real, el patrón Creador sugiere a *OperadorPlaca* como candidata para la creación de Mantención. La instancia *OperadorPlaca* podría enviar entonces el mensaje *añadir Tareas*, *añadir Estados*, *añadir Placa* a mantención, pasando la nueva Tarea, Estado y Placa como parámetros. La figura 6 muestra un primer diagrama de secuencia que refleja esto.

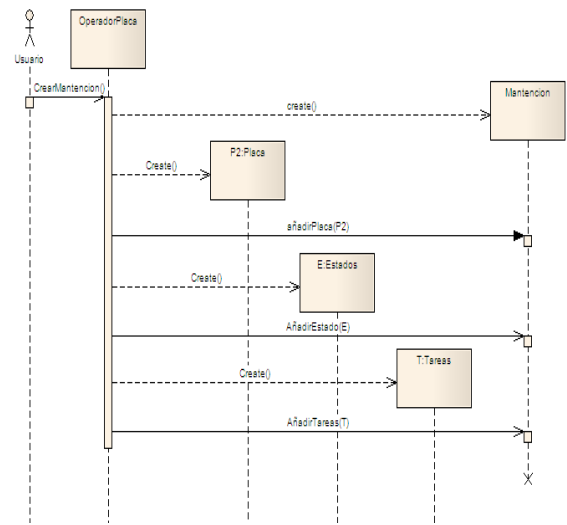

Figura 6. Operador Placa crea Placa, Estados y Tareas

Esta asignación de responsabilidades acopla la clase *OperadorPlaca* con el conocimiento de la clase Placa, Estados y Tareas.

Otra alternativa de diseño sería que la clase Mantención se encargue de crear las instancias Placa, Estados y

Tareas, tal como lo muestra la figura 7, que describe esta alternativa.

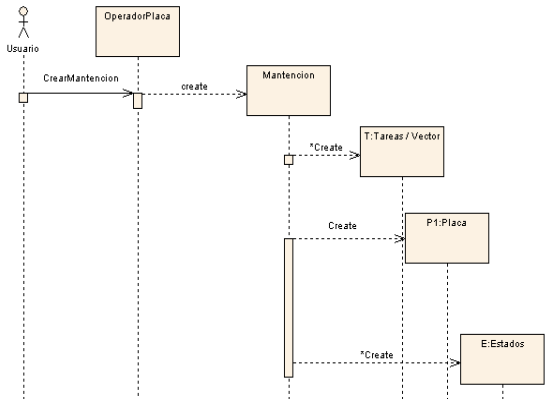

Figura 7. Mantención crea Placa, Estados y Tareas

¿Qué diseño, basado en la asignación de responsabilidades, soporta bajo acoplamiento?. En ambos casos se asume que la Mantención debe acoplarse con el conocimiento de Placa, Tareas y Estados. El diseño uno, en el que OperadorPlaca crea Placa, Estado y Tareas, añade acoplamiento entre OperadorPlaca y Placa, Estado y Tareas, mientras que el diseño dos, en el que Mantención crea estas instancias, no incrementa el acoplamiento. Desde el punto de vista puramente del acoplamiento, es preferible el diseño dos porque mantiene el acoplamiento global más bajo.

## DISEÑO DE SIGEP CON PATRONES GoF

El siguiente problema de diseño a resolver es proporcionar una lógica más compleja para crear una mantención, como crear una mantención de una placa de cátodo de cobre siempre y cuando esta placa no exceda el límite máximo de costos de mantención (punto crítico) o ya tenga una mantención en la misma fecha.

Es probable que más adelante puedan surgir nuevas restricciones al proceso de mantención como por ejemplo, no realizar mantención a placas de cátodo de cobre con estado pandeado[3] u otras variaciones.

Debido a esto, nos plateamos la siguiente cuestión ¿Cómo diseñar los diversos algoritmos de restricciones de mantención?

Una alternativa fue usar el Patrón GoF Estrategia [4], puesto que la restricción de la mantención varía según la estrategia (o algoritmo), se crean múltiples clases *EstrategiaCriterio*, cada uno con un método polimórfico *getCriterio*. A cada método *getCriterio* se le pasa como parámetro el objeto Placa, de manera que el objeto de la estrategia de restricción de mantención pueda encontrar el algoritmo adecuado. La implementación de cada *getCriterio* será diferente.

---
[3] Placa pandeada hace referencia a las placas que se doblan fácilmente al ejercer presión sobre ellas.

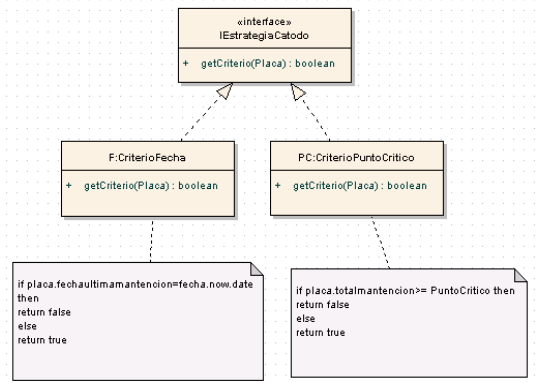

Figura 8. Patrón Estrategia

Un objeto estrategia se conecta a un objeto de contexto, el objeto al que se aplica el algoritmo. En este caso, el objeto de contexto es una Placa.

Existen diferentes algoritmos o estrategias de mantención, y cambian con el tiempo, aquí surge la interrogante ¿quién debería crear la estrategia?. Un enfoque directo es aplicar el patrón Factoría. La clase Factoría Singleton, contiene como parámetros de creación el tipo de estrategia a utilizar. La figura 9 muestra esta alternativa de diseño.

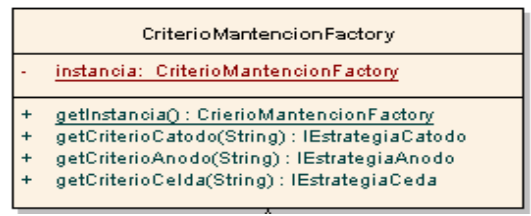

Figura 9. Factoría Singleton Mantención

El patrón Estrategia nos lleva al siguiente problema. ¿Cómo gestionar el caso de varias políticas de restricciones de mantención?. Por ejemplo aplicar restricción de fecha y punto crítico a la vez, ó las placas con estado pandeado pueden tener más de una mantención diaria, ó las placas con cierto estado no pueden tener mantención, etc.

¿Hay alguna forma de cambiar el diseño de manera que el objeto Placa no conozca si está tratando con una o más estrategias, y ofrecer también un diseño para la resolución de políticas contradictorias?.

Una buena solución a este problema es agregar al patrón GoF Estrategia, el patrón GoF Composite.

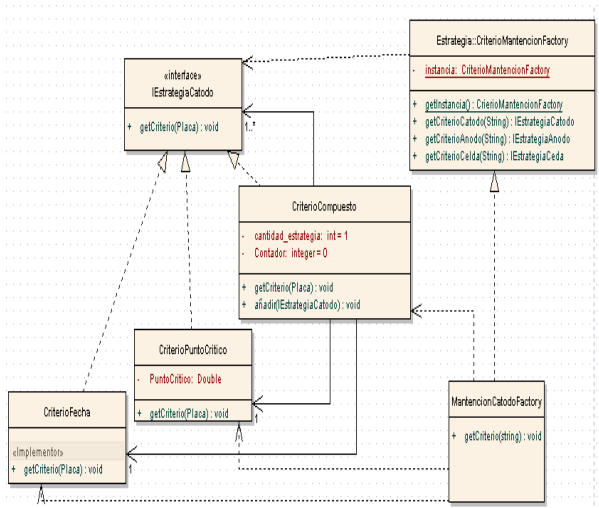

Figura 10. Patrón Estrategia, Composite, Factory y Singleton.

Con el patrón Composite, se ha creado un grupo de diferentes estrategias de restricción de mantención, que para Placa aparecen como una única estrategia.

En la etapa de requerimientos se definieron sólo dos restricciones de mantención de placas de cátodo de cobre, con este diseño se busca que la implementación de nuevos criterios de mantención a distintos elementos del sistema sea de forma fácil y poco traumática para los demás componentes del sistema.

## DISEÑO DEL FRAMEWORK

El software SIGEP, requiere que se almacene y recupere la información en mecanismos de almacenamiento persistente, como una base de datos relacional. Por tanto se necesita un servicio de persistencia que se construya con un FrameWork de Persistencia (FWP). Este es un framework simplificado que deberá proporcionar funciones para almacenar y recuperar los objetos en un mecanismo de almacenamiento persistente.

Las características esenciales del diseño de los conversores de base de datos, que constituyen una parte central del FWP, se basan en el patrón de diseño GoF Template Method (método plantilla). Este patrón es una parte esencial del diseño del framework, de manera más específica, los frameworks de caja blanca. Normalmente éstos son frameworks orientados a la definición de subclases y jerarquías de clases, que requieren que los usuarios conozcan algo acerca de su diseño y estructura, de ahí lo de caja blanca [7].

La idea es crear un método (método plantilla) en una superclase que define el esqueleto de un algoritmo, en sus partes variables e invariables. El Método Plantilla invoca otros métodos, algunos de los cuales podrían redefinirse en una subclase.

El punto de variación es la manera de crear el objeto a partir del almacenamiento. El método get será el método plantilla en una superclase abstracta ConversorPersistenciaAbstracto que define la plantilla, y utiliza un método "de enganche" en las subclases para la parte que varía, como se ilustra en la figura 11:

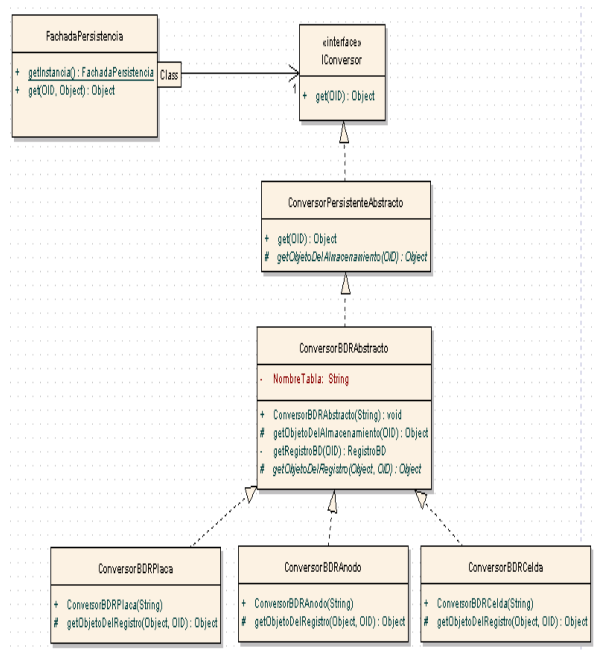

Figura 11. Conversor de datos con el Método Plantilla.

## ARQUITECTURA LÓGICA CON PATRONES

El sistema SIGEP está compuesto de muchos paquetes lógicos, como un paquete de interfaz de usuario, un paquete de acceso a base de datos, etc. Cada paquete agrupa un conjunto cohesivo de responsabilidades. Esta es la práctica básica de aplicar la modularidad para dar soporte a la separación de intereses.

El patrón Capas se relaciona con la arquitectura lógica, es decir, describe la organización conceptual de los elementos del diseño en grupos, independiente de su empaquetamiento o despliegue físico.

Las capas definen un modelo general de N-niveles para la arquitectura lógica, que produce una arquitectura en capas. Basado en el arquetipo de las capas comunes en una arquitectura lógica de un sistema de información, se ilustra en la figura 12 la arquitectura lógica en capas parcial de la aplicación SIGEP.

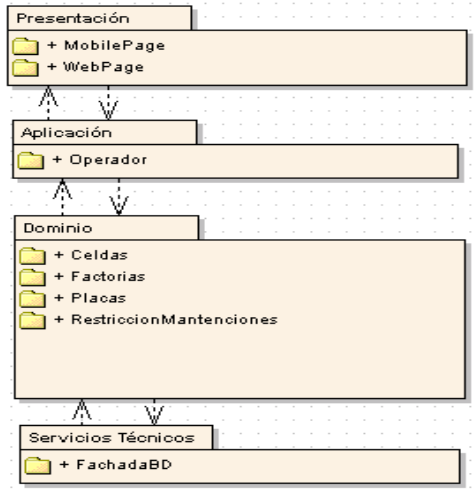

Figura 12. Vista lógica de las capas del sistema SIGEP.

El principio de Separación Modelo-Vista, es clave en el patrón Modelo-Vista-Controlador (MVC). Donde el Modelo es la Capa de Dominio, la Vista es la Capa de Presentación, y el Controlador son los objetos del flujo de trabajo en la Capa de Aplicación, que establece que los objetos del dominio no deberían conocer directamente a los objetos de la vista o presentación.

Este principio mantiene que las clases del dominio encapsulan la información y el comportamiento relacionado con la lógica de la aplicación. Las clases de las ventanas son relativamente delgadas, responsables de la entrada y salida, y capturan los eventos del UI (User Interface), pero no mantienen datos ni proporcionan directamente ninguna funcionalidad de la aplicación.

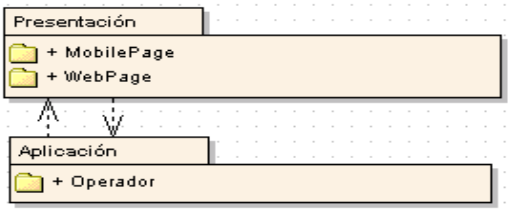

Figura 13. Separación Modelo-Vista.

En el sistema SIGEP las ventanas envían mensajes a la clase operador, consultando sobre la información que luego mostrarán como elementos gráficos, manteniendo así este principio.

## INTEGRACIÓN DE HERRAMIENTAS MÓVILES

Para la implementación de SIGEP, se utilizó la plataforma Visual Studio.NET 2003. Este ambiente de desarrollo, tiene un completo conjunto de herramientas para construir aplicaciones Web en ASP, Servicios Web XML, aplicaciones para computadores personales y aplicaciones para dispositivos móviles.

La aplicación móvil se implementó en dispositivos Intermec CK61 (figura 14). Estos equipos móviles son para uso industrial, son terminales recolectores de datos intermec, con gran potencia, fuerza y velocidad: El Terminal portátil CK61 de Intermec cumple todas las exigencias de los entornos más duros, es decir para condiciones climáticas y ambientales más exigentes. La batería tiene capacidad para funcionar durante todo el día y el diseño robusto del Terminal proporciona un funcionamiento fiable durante años. El procesador y el sistema operativo de nueva generación permiten ejecutar aplicaciones complejas. El almacenamiento no volátil protege la información importante.

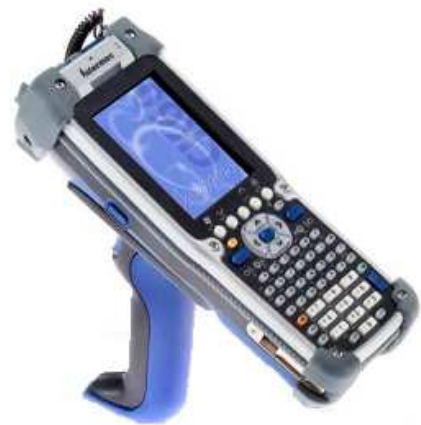

Figura 14. Equipo Intermec CK61

Estos equipos se configuraron y utilizaron para la toma de datos en línea, dentro de la faena de las mantenciones de las placas catódicas, en este equipo los operadores ingresan los datos de identificación de la placa, empresa que realiza la mantención, el estado en el que llegó la placa y las tareas de mantenciones a realizar (figura 15).

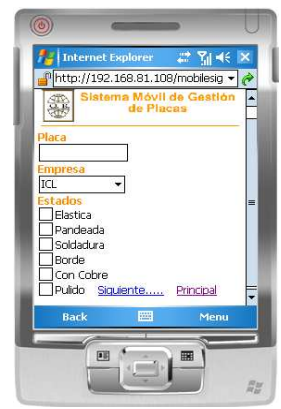

Figura 15. Pantalla ingreso de mantenciones

Para el acceso móvil al sistema, se implementó una red inalámbrica en el taller de reparaciones de las placas de cátodos de cobre, a través de la cual se integran los equipos clientes móviles Ck61 Intermec.

La figura 16 muestra la disposición de las particiones físicas del sistema SIGEP y la asignación de los componentes software a estas particiones, lo cual también permite visualizar la integración de las compenentes móviles al software SIGEP.

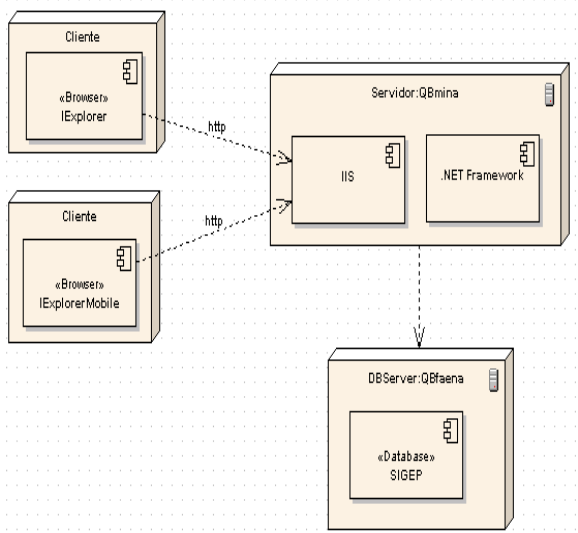

Figura 16. Integración de las componentes móviles al SIGEP.

## ANALISIS DE RESULTADOS

Desde el punto de vista operacional de gestión, la información muy útil que proporciona SIGEP, es el detalle del historial de las mantenciones de las placas catódicas y conocer cuáles de ellas genera mayores costos para la compañía como se aprecia en la figura 17.

Con SIGEP y su operación en línea del control de las mantenciones, se obtuvo el manejo exhaustivo de todos los datos que se generan en este proceso.

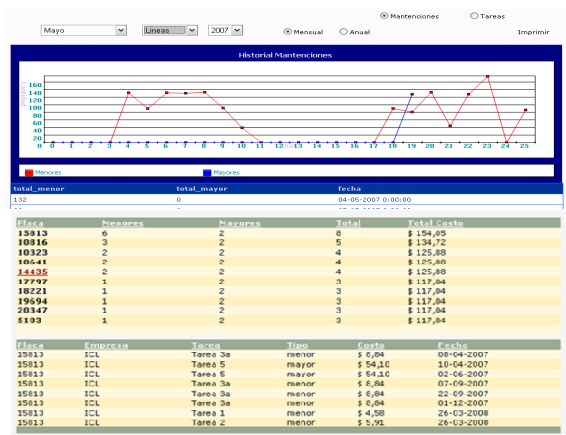

Figura 17. Listado de placas con mayor costo de mantenciones.

Desde punto de vista y objetivo de la reducción de los costos, para la empresa, en la operación de seis meses del software la empresa logró reducir sus costos de mantención en comparación con el periodo anterior en el mismo semestre, en cerca de 15%, debido a que el sistema arroja información que permite anticipadamente tomar decisiones acerca del recambio de las placas con mayor deterioro, o seguir una apropiada mantención de estas.

Respecto al desempeño técnico del sistema, se observó un rendimiento óptimo en cuanto a la comunicación y tiempos de respuestas, entre los clientes fijos y móviles, respecto a las interacciones con el servidor y la base de datos, aunque desde un punto de vista solamente cualitativo, porque no se hicieron mediciones al respecto por lo que datos cuantitativos sobre este aspecto no es posible entregar.

## CONCLUSIONES

En el desarrollo de este trabajo, se persiguió dar solución a las problemáticas puntuales que tiene la superintendencia de planta de CMQB S.A., respecto a las mantenciones de las placas de cátodos de cobre, en lo que se refiere a llevar un adecuado control y seguimiento de las tareas realizadas a estos elementos y así evitar posibles cobros y costos excesivos debido a la desinformación existente.

En este aspecto SIGEP cumple con el objetivo de entregar información relevante que permite llevar un adecuado control de las tareas realizadas a las placas catódicas, lo que ha permitido hasta el presente, disminuir las tareas realizadas a las placas de cátodo de cobre hasta el punto de reducir en un turno los trabajos realizados por este concepto.

La utilización de patrones de diseño en SIGEP permitió establecer y otorgarle flexibilidad en el diseño, independiente y extensible en el tiempo, al separar la interfaz y el diseño, esta separación hizo más fácil la creación y reutilización de código.

Además, al utilizar patrones se tornó más fácil documentar los detalles del diseño y como reutilizar la aplicación. Así también, se comprobó que son diseños muy efectivos y eficientes, ampliamente demostrados por la experiencia de otras personas y que ayudan a construir software correctamente.

De esta manera, se logró el objetivo de crear una aplicación de fácil entendimiento, modificación o adaptación del diseño por terceras personas, lo cual fue un requerimiento específico de la superintendencia de planta de la CMQB S.A.

El framework construido para SIGEP, es una solución breve a los problemas y soluciones de diseño de un framework y servicios de persistencia. Se han encubierto muchas puntos importantes como: seguridad, acceso multiusuario, estrategias de bloqueo, y gestión de errores.

Para el presente año, se contempla iniciar la renovación de unas cinco mil placas de cátodos para la CMQB S.A., por lo que el sistema SIGEP se convierte en una poderosa herramienta para decidir cuales placas de cátodos son las que deben ser reemplazadas de acuerdo a sus costos de mantención.

## REFERENCIAS

[1] G. Booch, J. Rumbaugh, y I. Jacobson. El Lenguaje Unificado de Modelado. Madrid: ddison Wesley Iberoamericana, 1999.

[2] W. R. Greiff. "Paradigma vs Metodología; El caso de la POO (Parte II)". Soluciones Avanzadas, 1994.

[3] L. Bass, P. Clements y R. Kazman, "Software Architecture in Practice", Addison- Wesley, 1998.

[4] E. Gamma, R. Helm, R. Johnson y J. Vlissides, "Design Patterns: Elements of Reusable Object-Oriented Software" (Gang of Four, [GoF]). Addison-Wesley Professional Computing Series. [GoF95], 1995.

[5] C. Alexander, S. Murria, M. Jacobson, I. Fiksdahl-King, y S. Angel, "A pattern Language: Towns/Building/Construction". Oxford University Press, 1977.

[6] C. Larman, "UML y Patrones, Una introducción al análisis y diseño orientado a objetos y al proceso unificado"; Prentice Hall, 2002.

[7] M. Fowler, "Patterns of Enterprise Application Architecture", Adison Wesley, 2003.